\documentclass[preprint,12pt]{elsarticle}

\usepackage{amssymb}
\usepackage{amsmath}
\usepackage{array}
\usepackage{epstopdf}
\usepackage[mathscr]{euscript}

\journal{Superlattices and Microstructures}

\begin{document}

\begin{frontmatter}

\title{Crossing and anti-crossing effects of polaritons in a~magnetic-semiconductor superlattice influenced by an~external magnetic field}

\author[cent,rian,opto]{Vladimir~R.~Tuz}
\author[rian]{Volodymyr~I.~Fesenko\corref{correspond}}
\ead{v.i.fesenko@ieee.org}
\author[poly]{Illia~V.~Fedorin}
\author[opto]{Hong-Bo~Sun}
\author[cent,rian,phys]{Valeriy~M.~Shulga}

\address[cent]{International Center of Future Science, Jilin University, Changchun, \\ People's Republic of China}
\address[rian]{Institute of Radio Astronomy of National Academy of Sciences of Ukraine, \\ Kharkiv, Ukraine}
\address[opto]{State Key Laboratory on Integrated Optoelectronics, College of Electronic Science and Engineering, Jilin University, Changchun, People's Republic of China}
\address[poly]{National Technical University `Kharkiv Polytechnical Institute', Kharkiv, Ukraine}
\address[phys]{College of Physics, Jilin University, Changchun, People's Republic of China}

\cortext[correspond]{Institute of Radio Astronomy of National Academy of Sciences of Ukraine, 4, Mystetstv St., Kharkiv 61002, Ukraine}

\begin{abstract}
Crossing and anti-crossing effects in dispersion characteristics of both bulk and surface polaritons in a magnetic-semiconductor superlattice influenced by an external static magnetic field being in the Faraday geometry are discussed. The bulk polaritons are classified as eigenwaves with right-handed and left-handed elliptically polarized states, whereas the surface polaritons are considered as hybrid modes having a predominant effect of either magnetic or semiconductor subsystem, and distinctions in dispersion characteristics of such polaritons are revealed involving the concept of critical points. 
\end{abstract}


\end{frontmatter}



\section{Introduction}
\label{sec:intro}

Effects of crossing, avoided crossing (anti-crossing), splitting and coalescence of modes have significant impact on functionality and efficiency of many microwave and optical devices and, therefore, they are a subject of intense studies over decades (see \cite{Pfeiffer_PhysRevB_1974, Mertens_RadioSci_1996, Shestopalov_Electromagn_1993, Yakovlev_MTT_1998, Yakovlev_MTT_2000,Sauer_PPCF_2012, Torma_RPP_2015, Dadoenkova_SL_2016, Refki_Plasm_2016,Wilkes_SL_2017} and references therein). These effects imply that several modes coexist in the same frequency band and wavevector space, moreover, their dispersion curves appear to be closely spaced, and in the band of their convergence the curves can intersect, merge each other or diverge apart. Remarkably, when the dispersion curves of such modes intersect each other exhibiting the \textit{crossing} effect, the corresponding waves are not coupled or coupling between them is negligible small. It means that the nature of each mode remains unchanged, and, thus, the modes just degenerate at the frequency of their crossing point. In contrast to the crossing effect, the \textit{anti-crossing} effect is related to the phenomenon of strong modes coupling which can result in the modes type transformation and/or energy exchange between the waves. Such modes type transformation can appear between waves of the same or different nature, and, in particular, the mutual inter-type conversion between surface, guided and bulk modes can occur. Moreover, these waves can possess different conditions being propagating, leakage or evanescent ones. The anti-crossing effect emerges in dispersion curves of corresponding modes in the form of a significant variation of their slope within the convergence band where the curves become to be bent before and after the crossing point, exhibiting either normal or anomalous dispersion line. In particular, it leads to a sharp change in the group velocity of the coupled waves and, therefore, from this viewpoint the anti-crossing effect is beneficial for practical applications.

Indeed, the utilization of the modes crossing and anti-crossing effects is widely discussed for different types of waveguide systems and resonators. Thus, it is revealed \cite{Pfeiffer_PhysRevB_1974, Novotny_PhysRevE_1994, PengFei_ChinPhysB_2008} that the surface modes existing in waveguides and optical fibers having an imperfectly conducting cladding (i.e. optical characteristics of such a cladding are described by a complex dielectric function that accounts both material dispersion and losses) can be transformed into waveguide modes and vice versa, due to the anti-crossing effect. It is found out that in such waveguide systems above the plasmon frequency surface modes convert into evanescent modes, whereas surface and guided modes can either couple between each other or transform into each other. The latter occurs when in addition to the phase constants, the attenuations of both modes also come close. Further, it is shown that in much more sophisticated waveguide systems consisting of an anisotropic composite filling \cite{Neira_SciRep_2015} the anti-crossing effect between bulk and guided modes can lead to appearance of conditions for slow, stopped or superluminal propagation of waves accompanied by their very strong group velocity dispersion. It is proposed to use these modes for designing stopped-light nanolasers for nanophotonic applications and dispersion-facilitated signal reshaping in telecommunications. Besides, a manifestation of both crossing and anti-crossing effects for surface (plasmon-polariton and exciton-polariton) modes in the systems comprising a graphene sheet and a quantum well structure embedded into a resonant microcavity was recently reported in \cite{Zhao_ApplPhys_2015} and \cite{Gao_2015}, respectively. 

Remarkable, a manifestation of the crossing and anti-crossing effects in such different systems can be considered within a unified theory. As the simplest one, a coupled harmonic oscillator model should be mentioned \cite{Novotny_AmJPhys_2010} which allows to understand the effects at an intuitive level. In the framework of this model the mode-coupling factor \cite{Pierce_JApplPhys_1954} is considered as a special parameter, which is intended to characterize the modes coupling strength. Besides, more general and formalized approach is developed, which is based on the oscillation theory with involving the concept of equilibrium points from the catastrophe and bifurcation theories \cite{Gilmore_book_1981}, in order to describe the mutual mode coupling via derivation of the Morse critical points. Such an approach has been applied for analysis of a variety of waveguide and resonator structures \cite{Shestopalov_Electromagn_1993, Yakovlev_MTT_1998, Yakovlev_MTT_2000} that demonstrates its efficiency for careful identification of mode-coupling regions, and for reconstruction of dispersion behavior in those regions via simple analytical expressions.   
 
In our previous publications \cite{Fesenko_OptLett_2016, Tuz_JMMM_2016, Tuz_arXiv_2016} it is demonstrated that the dispersion features of both bulk and surface polaritons that propagate through a magnetic-semiconductor superlattice can be rather complicated. Generally, it is found out that in such a structure being under an action of the external static magnetic field the bulk polaritons appear as ordinary and extraordinary elliptically polarized waves with very different characteristics, whereas the surface polaritons are of hybrid type that can possess branches of anomalous dispersion. Moreover, there are particular bands where dispersion curves of these waves exhibit either crossing or anti-crossing effect. 

Therefore, in this paper it is our goal with involving procedures of the analytical theory about the Morse critical points \cite{Shestopalov_Electromagn_1993, Yakovlev_MTT_1998} to reveal peculiarities of manifestation of the crossing and anti-crossing effects in a magnetic-semiconductor superlattice influenced by an external static magnetic field being in the Faraday geometry, and to demonstrate that the conditions of appearance of these effects can be achieved by providing a proper choice of both geometric and material parameters of the structure under study.

\section{Superlattice description and dispersion relations}
\label{sec:dispersion}

\begin{figure}[htbp]
\centerline{\includegraphics[width=0.7\columnwidth]{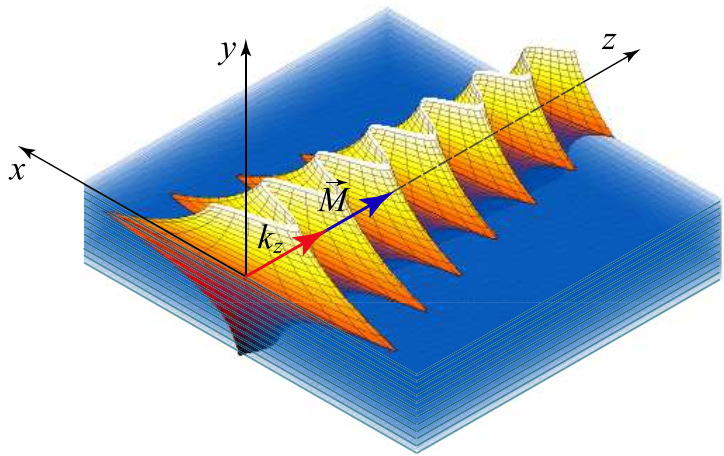}}
\caption{The problem sketch with a visual representation of the tangential electric field distribution ($E_x$) of the surface polariton which propagates through a magnetic-semiconductor superlattice influenced by an external static magnetic field being in the Faraday geometry.} \label{fig:fig_SPP}
\end{figure}

Thereby, further in this paper we study polaritons dispersion features of a semi-infinite stack (superlattice) of identical double-layer slabs (unit cells) that are periodically arranged along the negative direction of the $y$-axis (Fig.~\ref{fig:fig_SPP}). The structure unit cell consists of two constitutive layers made of magnetic (with constitutive parameters $\varepsilon_m$, $\hat \mu_m$) and semiconductor (with constitutive parameters $\hat \varepsilon_s$, $\mu_s$) materials having thicknesses $d_m$ and $d_s$, respectively, and, thus, the structure period is $L = d_m + d_s$. We also assume that the superlattice is infinite along the $x$ and $z$ directions and it adjoins an isotropic medium with constitutive parameters $\varepsilon_0$, $\mu_0$ occupying upper half-space $y>0$. 

It is supposed that the structure under study is influenced by an external static magnetic field  $\vec M$ directed along the $z$-axis, and the strength of this magnetic field is high enough to form a homogeneous saturated state in both magnetic and semiconductor constitutive layers. We designate that electromagnetic waves propagate along the $z$-axis, i.e. the wavevector $\vec k$ of surface waves has components $\{0,\pm i\kappa,k_z\}$, where $\kappa$ is responsible for the wave attenuation in the positive ($+i\kappa$) and negative ($-i\kappa$) directions of the $y$-axis. In such a problem statement the directions of vectors $\vec M$ and  $\vec k$ are coincident, therefore, the structure is considered to be in the \textit{Faraday} geometry. 

Hereinafter we assume that all characteristic dimensions (the layers thicknesses and period length) of such multilayered structure are much smaller than the wavelength in the corresponding part of the superlattice (the long-wavelength limit, $d_m\ll \lambda$, $d_s \ll \lambda$, $L \ll \lambda$), i.e. the structure is a finely-stratified one. With taking into account the long-wavelength limit and involving averaging (homogenization) procedures from the effective medium theory (is not presented here; for details, see, \cite{Tuz_JMMM_2016, Tuz_arXiv_2016, Agranovich_SolidStateCommun_1991, Elmzughi_JPhysCondMat_1995}), the superlattice is further described by two tensors of relative effective permittivity and relative effective permeability written in the form:
\begin{equation}
 \hat \varepsilon=\left( {\begin{matrix}
   {\varepsilon_{xx}} & {\varepsilon_{xy}} & 0 \cr
   {-\varepsilon_{xy}} & {\varepsilon_{yy}} & 0 \cr
   0 & 0 & {\varepsilon_{zz}} \cr
\end{matrix}
} \right),~
\hat\mu=\left( {\begin{matrix}
   {\mu_{xx}} & {\mu_{xy}} & 0  \cr
   {-\mu_{xy}} & {\mu_{yy}} & 0  \cr
   0 & 0 & {\mu_{zz}}  \cr
 \end{matrix}
} \right). \label{eq:eff}
\end{equation}
Under such an approximation, the problem is reduced to characterization of the magnetic-semiconductor multilayered structure by an equivalent biaxial anisotropic (\textit{gyroelectromagnetic}) uniform medium, whose first optical axis is directed along the structure periodicity, while the second one coincides with the direction of the external static magnetic field $\vec M$, i.e., in general, for this gyroelectromagnetic medium the next conditions are met: $\varepsilon_{xx} \neq \varepsilon_{yy} \neq \varepsilon_{zz}$ and $\mu_{xx} \neq \mu_{yy} \neq \mu_{zz}$. Expressions for components of effective constitutive tensors \eqref{eq:eff} derived via parameters of underlying magnetic ($\varepsilon_m$, $\hat \mu_m$) and semiconductor ($\hat \varepsilon_s$, $\mu_s$) layers as well as their dispersion characteristics are omitted here and can found in our previous publications \cite{Tuz_PIERB_2012, Tuz_JO_2015, Tuz_Springer_2016}. In particular, these parameters are calculated in the microwave part of spectrum for a magnetic-semiconductor structure made in the form of a barium-cobalt/doped-silicon superlattice \cite{Wu_JPhysCondensMatter_2007}. Remarkably, for such a structure the characteristic resonant frequencies of underlying constitutive magnetic and semiconductor materials appear to be different but rather closely spaced within the same frequency band.

In order to derive a solution with respect to bulk and surface polaritons the approach developed in \cite{Burstein_PhysRevB_1974, Abraha_1995} is extended to the case of a gyroelectromagnetic medium, whose relative permeability as well as relative permittivity simultaneously are tensor quantities (see \ref{sec:dispeq}; for a general treatment also see \cite{borisov_OptSpectr_1994, Tsurumi_JPSJ_2007, Lakhtakia_book_2013}). It gives us two dispersion relations $\mathscr{D}(k_z,k_0)=0$ written in the form
\begin{subequations}
\label{eq:DispEq}
\begin{align}
\label{eq:DispBulk}
&\mbox{\ae}_x^2\mbox{\ae}_y^2 - k_0^4\varsigma_{xy}\varsigma_{yx}=0,\\
\label{eq:DispSurf}
\begin{split}
(&\kappa_{2} +\kappa_{0} g_{zz})(\kappa_{2}^2 \varsigma_{yy}-\mbox{\ae}_y^2\varsigma_{zz}) \Bigl\{\kappa_{1}\zeta(\kappa_{0}^2-k_z^2)\\& +\kappa_{0}\left[\kappa_{1}^2 g_{xy}\varsigma_{yy}-\varsigma_{zz}(k_0^2 \xi + k_z^2 g_{xy})\right] \Bigr\}\\
-(&\kappa_{1}+\kappa_{0} g_{zz})(\kappa_{1}^2 \varsigma_{yy}-\mbox{\ae}_y^2\varsigma_{zz})\Bigl\{\kappa_{2}\zeta(\kappa_{0}^2-k_z^2) \\& +\kappa_{0}\left[\kappa_{2}^2g_{xy}\varsigma_{yy} -\varsigma_{zz}(k_0^2 \xi + k_z^2 g_{xy})\right]\Bigr\}=0,
\end{split}
\end{align}
\end{subequations}
for bulk and surface polaritons, respectively. Here $\varsigma_{\nu\nu'}$ are elements of the tensor $\hat \varsigma$ which is introduced as the product of tensors $\hat \mu$ and $\hat \varepsilon$ made in the appropriate order (in what follows subscripts $\nu$ and $\nu'$ are substituted to iterate over corresponding indexes of the tensor quantities in Cartesian coordinates); for surface polaritons two distinct substitutions $\varepsilon_{\nu\nu'} \to g_{\nu\nu'}$, $\varepsilon_v \to g_v$ and $\mu_{\nu\nu'} \to g_{\nu\nu'}$, $\mu_v \to g_v$ correspond to the problem resolving with respect to vectors $\vec E$ and $\vec H$, respectively (here we kindly ask the reader to compare the solution procedures given in \cite{Burstein_PhysRevB_1974} and \cite{Abraha_1995} for gyroelectric (semiconductor) and gyromagnetic (antiferromagnetic) superlattices, respectively); $\zeta=g_{v}g_{yy}\varsigma_{yx}$, $\xi=g_{yy}\varsigma_{yx}-g_{xy}\varsigma_{yy}$, $\mbox{\ae}_\nu^2=k_z^2-k_0^2\varsigma_{\nu\nu}$, and $\varepsilon_v=\varepsilon_{xx} + \varepsilon_{xy}^2/\varepsilon_{yy}$ and $\mu_v=\mu_{xx} + \mu_{xy}^2/\mu_{yy}$ can be considered as relative transverse permittivity and permeability, respectively. The remaining notations are given in \ref{sec:dispeq}.

Notice, in two particular cases of a medium which is characterized by either scalar permeability $\mu$ and tensor permittivity $\hat\varepsilon$ or tensor permeability $\hat\mu$ and scalar permittivity $\varepsilon$, dispersion relation \eqref{eq:DispSurf} for surface polaritons coincides with Eq.~(38) of \cite{Burstein_PhysRevB_1974} and Eq.~(13) of \cite{Abraha_1995} for gyroelectric and gyromagnetic superlattices, respectively, that verifies the obtained solution.

\section{Crossing and anti-crossing effects}
\label{sec:effects}

\begin{figure}[htbp]
\centerline{\includegraphics[width=0.7\columnwidth]{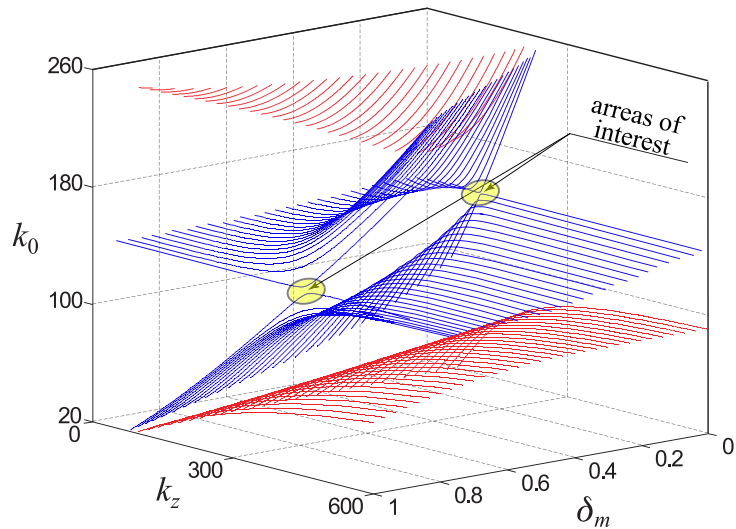}}
\caption{A set of dispersion curves  of right-handed (red curves) and left-handed (blue curves) elliptically polarized bulk polaritons for different filling factor $\delta$ of a magnetic-semiconductor superlattice.} \label{fig:fig_BulkSet}
\end{figure}

\begin{figure}[htbp]
\centerline{\includegraphics[width=0.45\columnwidth]{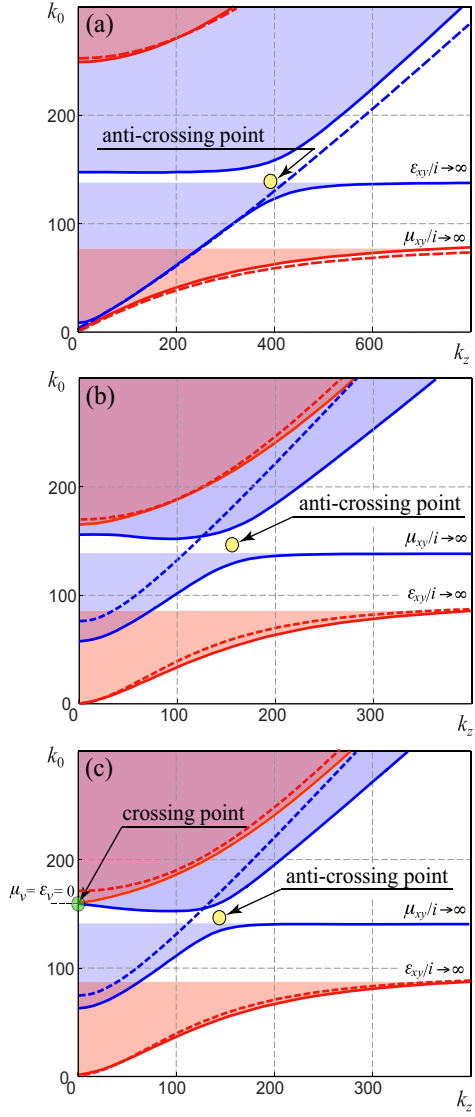}}
\caption{Solid lines correspond to dispersion curves that outline passbands of right-handed (red) and left-handed (blue) \textit{elliptically} polarized bulk polaritons of a superlattice with a predominant impact either magnetic (a) or semiconductor (b, c) subsystem; dashed lines correspond to dispersion curves of right-handed (red) and left-handed (blue) \textit{circularly} polarized bulk polaritons of reference magnetic (a) and semiconductor (b, c) homogeneous media; (a)~intermediate interaction, $\delta_m=0.9$; (b)~strong interaction, $\delta_s=0.85$; (c)~accidental degeneracy, $\delta_s=0.9$.} \label{fig:fig_Bulk}
\end{figure}

\begin{figure}[htbp]
\centerline{\includegraphics[width=0.6\columnwidth]{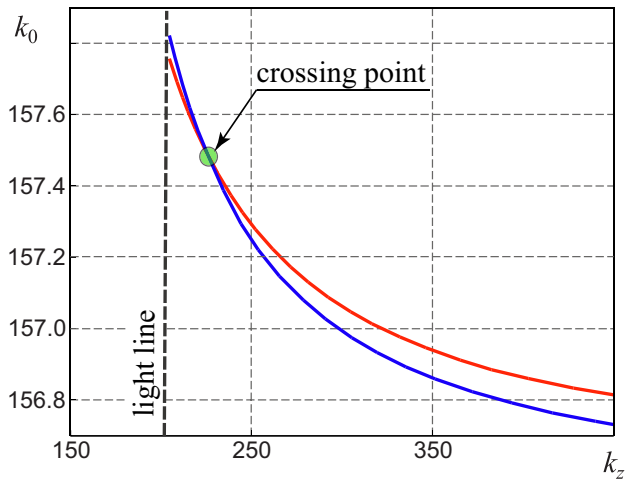}}
\caption{Manifestation of the crossing effect in dispersion curves of the hybrid EH (red line) and HE (blue line) surface polaritons; $\delta_m=0.15$.} \label{fig:fig_Surf}
\end{figure}

In order to obtain dispersion characteristics of the magnetic-semiconductor superlattice under study dispersion equations \eqref{eq:DispEq} are solved numerically. The results of our calculations are summarized in Figs.~\ref{fig:fig_BulkSet}, \ref{fig:fig_Bulk} and \ref{fig:fig_Surf} for the bulk and surface polaritons, respectively.

A complete set of dispersion curves that outline the bands of existence (passbands) of the bulk polaritons as a function of the filling factor $\delta$ ($\delta_{m}=d_m/L$, $\delta_{s}=d_s/L$,  $\delta_{m}+\delta_{s}=1$) is presented in Fig.~\ref{fig:fig_BulkSet}. In what follows we define that the structure has a configuration with a \textit{predominant} impact of the magnetic subsystem if $\delta_{m}\gg\delta_{s}$, whereas it has that one of the semiconductor subsystem if $\delta_{s}\gg\delta_{m}$. Moreover, since the given superlattice is characterized by the biaxial anisotropy, for which as already mentioned the first anisotropy axis is associated with the structure periodicity while the second one is a result of the external static magnetic field influence, the eigenwaves (bulk polaritons) of the equivalent unbounded gyroelectromagnetic medium have right-handed and left-handed \textit{elliptical} polarization. We distinguish this different polarization of the bulk polaritons in their dispersion curves in Fig.~\ref{fig:fig_BulkSet} by red and blue colors, respectively. One can conclude that for the bulk polaritons of each polarization there is a pair of corresponding sets of dispersion curves separated by a forbidden band (stopband). We are mainly interested in those curves of the sets which have greatly sloping branches and exhibit the closest approaching each other, since such dispersion behaviors are a direct manifestation of the crossing and anti-crossing effects. In particular, two areas of interest are distinguished in Fig.~\ref{fig:fig_BulkSet} by yellow circles. It is found out that convergence of dispersion curves is only peculiar for the left-handed elliptically polarized waves propagating through the structure whose filling factor $\delta$ corresponds to either magnetic or semiconductor predominant subsystem.

Based on the identified filling factor $\delta$ of the structure under study three particular $k_z-k_0$ planes are subsequently selected where there is a convergence of the dispersion curves. These planes are depicted in Fig.~\ref{fig:fig_Bulk} where, as previously, the dispersion curves of the bulk polaritons having right-handed and left-handed elliptical polarization are plotted with the red and blue solid lines, respectively, whereas  the colored areas designate  passbands of the corresponding waves. For a clear understanding of the physical picture of the discussed effects all dispersion curves of the bulk polaritons of the given superlattice are compared with those ones of the right-handed and left-handed \textit{circularly} polarized waves propagating through a reference unbounded gyromagnetic ($\delta_m=1$, $\delta_s=0$) or gyroelectric ($\delta_m=0$, $\delta_s=1$) homogeneous medium. Dispersion curves of these circularly polarized waves are plotted on the corresponding planes in Fig.~\ref{fig:fig_Bulk} with the red and blue dashed lines. 

One can conclude that for the chosen parameters of the structure under study the  dispersion curves of the right-handed elliptically polarized bulk polaritons appear to be quite trivial and they completely inherit characteristics of the right-handed circularly polarized waves of the corresponding reference medium. Contrariwise, the dispersion features of the left-handed elliptically polarized bulk polaritons are much more complicated being strongly dependent on the filling factor $\delta$ and resonant frequencies of constitutive parameters of the superlattice underlying materials. Therefore, in what follows we focus only on the consideration of dispersion features of the bulk polaritons having the left-handed polarization.

Indeed, in contrast to the characteristics of the left-handed circularly polarized waves of the corresponding reference medium whose passband has no discontinuity, the passband  of the left-handed elliptically polarized bulk polaritons of the superlattice is separated into two distinct areas. This separation is centered around the frequency at which the resonances of the functions $\varepsilon_{xy}/i$ and  $\mu_{xy}/i$ occur for the structure configurations with predominant impact of the magnetic and semiconductor subsystem, respectively. Remarkably, in both structure configurations the dispersion curves demonstrate significant variation of their slope having subsequent branches with normal and anomalous dispersion that possess approaching at some points (extreme states).  Also a particular extreme state in dispersion curves is found out where the branches of the left- and right-handed elliptically polarized bulk polaritons merge with each other.

From the mathematical point of view the found extreme states in dispersion curves of the bulk polaritons exist in the band where the differential $\mathscr{D}^{'}(k_z,k_0)$ of dispersion equation (\ref{eq:DispBulk}) vanishes. In order to identify the conditions of these extreme states an analytical approach based on the theory of Morse critical points \cite{Shestopalov_Electromagn_1993, Yakovlev_MTT_1998, Yakovlev_MTT_2000} can be involved. According to this theory the Morse critical points are generally defined by a set of nonlinear differential equations written in the form: 
\begin{subequations}
\label{eq:MCP}
\begin{align}
\label{eq:MCP1}
\mathscr{D}^{'}_{k_z}(k_z,k_0)\mid _{(k^m_z,k_0^m)} =\mathscr{D}^{'}_{k_0}(k_z,k_0)\mid _{(k^m_z,k_0^m)}&=0,\\
\label{eq:MCP2}
\begin{split}
\mathbb{H} = \left[h_{11}h_{22}-h_{12}h_{21}\right] \mid _{(k^m_z,k_0^m)} & \neq 0,
\end{split}
\end{align}
\end{subequations}
where $(k^m_z,k_0^m)$ are coordinates in the $k_z-k_0$ plane of a particular $m$-th Morse critical point, the subscripts $k_z$ and $k_0$ near the letter $\mathscr{D}$ define corresponding partial derivatives $\partial/\partial k_z$ and $\partial/\partial k_0$, and $\mathbb{H}$ is the Hessian determinant with elements $h_{11}=\mathscr{D}^{''}_{k_z k_z}(k_z,k_0)$, $h_{12}=\mathscr{D}^{''}_{k_z k_0}(k_z,k_0)$, $h_{21}=\mathscr{D}^{''}_{k_0 k_z}(k_z,k_0)$ and $h_{22}=\mathscr{D}^{''}_{k_0 k_0}(k_z,k_0)$. From this set of equations each extreme state can be uniquely identified from the sign of the Hessian determinant \cite{Gilmore_book_1981}. Thus, when $\mathbb{H}<0$ the corresponding Morse critical point represents a saddle point which occurs in the region of a modal coupling (the anti-crossing effect), whereas in the case of degeneracy, when $\mathbb{H}=0$, it is a non-isolated critical point (the crossing effect). In this way all found extreme states are classified and marked in Fig.~\ref{fig:fig_Bulk} by green and yellow circles for the crossing and anti-crossing effects, respectively. 

Moreover, in general, when conditions (\ref{eq:MCP}) are met, the type of interacting modes in the vicinity of the corresponding Morse critical point are defined as follows \cite{Yakovlev_MTT_1998}:
\begin{subequations}
\label{eq:Type}
\begin{align}
\label{eq:cod_forward}
\text{Codirectional forward:}~h_{12}/h_{11}<0,~ h_{22}/h_{11}>&0; \\
\label{eq:cod_backward}
\text{Codirectional backward:}~h_{12}/h_{11}>0,~ h_{22}/h_{11}>&0; \\
\label{eq:contr}
\text{Contradirectional:}~ h_{12}/h_{11}>0,~ h_{22}/h_{11}<&0.
\end{align}
\end{subequations}
For the problem under study condition \eqref{eq:contr} is met for all found extreme states. It means that along the converging branches on each side of the Morse critical point the bulk polaritons have different propagation conditions being either forward or backward waves (we consider a backward wave as a wave in which the direction of the Poynting vector is opposite to that of its phase velocity \cite{Lindell_MOP_2001}). Indeed, it is revealed that the waves are backward propagating on the corresponding branches where the anomalous dispersion takes place.

The strength of modes interaction within the found extreme states of the bulk polaritons can further be identified considering the classification introduced in \cite{Ibanescu_PhysRevLett_2004}. Thus, in the structure with a predominant impact of the magnetic subsystem the modes exhibit an \textit{intermediate interaction} (Fig.~\ref{fig:fig_Bulk}a) for which an appearance of flattened branches in dispersion curves is peculiar. In particular, a flattened branch arises before the anti-crossing point in the dispersion curve which outlines the upper passband. Besides, in the structure with a predominant impact of the semiconductor subsystem, the modes acquire a \textit{strong interaction} (Fig.~\ref{fig:fig_Bulk}b) which results in the branch formation having anomalous dispersion line. Within this branch the wave possesses backward propagation conditions. Finally, in the case of manifestation of the crossing effect, when two branches are merged within the critical point the modes are \textit{accidentally degenerate} (Fig.~\ref{fig:fig_Bulk}c).

We should note that in \cite{Ibanescu_PhysRevLett_2004} some unusual and counter-intuitive consequences of such behaviors of the dispersion curves (e.g. backward waves propagation, reversed Doppler shift, reversed Cherenkov radiation, atypical singularities in the density of states, etc.) for the TE and TM modes of an axially uniform waveguide are discussed, and it is emphasized that these effects are of considerable significance for practical applications.  

In order to elucidate the dispersion features of the surface polaritons the problem is decomposed into two particular solutions with respect to the vectors $\vec{H}$ and $\vec{E}$ \cite{Burstein_PhysRevB_1974, Abraha_1995}. The reason of such problem decomposition is that for the surface polaritons propagating over the interface of the magnetic-semiconductor superlattice under study the electromagnetic field has all six nonzero components, whereas two particular solutions give different relations between the magnitudes of  these components. It is specific for the hybrid waves \cite{Mertens_RadioSci_1996}, and, thus, the surface polaritons can be further classified as the hybrid EH (with $E_z > H_z$) and HE (with $H_z > E_z$) modes which follow from the solutions of the initial problem derived with respect to the vectors $\vec{H}$ and $\vec{E}$, respectively. 

For both initial problem considerations, dispersion equation (\ref{eq:DispSurf}) has four roots from which two physically correct ones ($\kappa_1$ and $\kappa_2$) must be selected \cite{Burstein_PhysRevB_1974} ensuring they correspond to the attenuating waves. In our study we consider only those  roots of equation (\ref{eq:DispSurf}) that are real and positive quantities (i.e. they correspond to the bonafide waves \cite{Burstein_PhysRevB_1974}).

Moreover, among all possible appearance of the surface polaritons dispersion curves we are interested in such branches which manifest the crossing or anti-crossing effect. The search for their existence implies solving an optimization problem, where for the crossing effect the degeneracy point must be found, whereas for the anti-crossing effect the critical points must be defined from calculation of the first $\mathscr{D}^{'}(k_z,k_0)$ and second $\mathscr{D}^{''}(k_z,k_0)$ partial derivatives of equation (\ref{eq:DispSurf}) with respect to the $k_z$ and $k_0$ for both EH and HE modes. During the solution of this problem the period and  constitutive parameters of the underlying materials of the superlattice are fixed, and the search for the effects manifestation is proceeded via altering the layers thicknesses within the period. The found  crossing effect is depicted in Fig.~\ref{fig:fig_Surf} for the structure with a predominant impact of the semiconductor subsystem ($\delta_m=0.15$), where the resolved degeneracy point for the hybrid EH (red line) and HE (blue line) surface polaritons is distinguished by a green circle. Remarkable, in this particular case  both dispersion curves possess an anomalous dispersion line. They start on the light line and fall just to the right of the light line, then they flatten out and approach an asymptotic limit for large values of $k_z$. 

It is revealed that the dispersion curves of the EH and HE surface polaritons exhibit the crossing effect exactly at the frequency at which $\varepsilon_{v}=0$, i.e. the surface polaritons appear to be completely degenerate at this point. Besides, the slopes of these dispersion curves do not coincide, and, therefore, at the crossing point the EH and HE surface polaritons have the same propagation constant $k_z$ whereas their group-velocities are different. 
 
\section{Conclusions}

To conclude, we have examined dispersion characteristics of both bulk and surface polaritons in a magnetic-semiconductor superlattice influenced by an external static magnetic field in the Faraday geometry. In these characteristics the crossing and anti-crossing effects are identified and investigated with an assistance of the analytical theory about the Morse critical points.

It is observed that in the case when specific conditions related to the constitutive parameters of the underlying materials and filling factor of the superlattice are satisfied, some unusual dispersion behaviors of the bulk polaritons appear, namely: (i) the anti-crossing effect with the gap formation between two branches of the left-handed elliptically polarized waves occurs, and (ii) the crossing effect between the high-frequency branches of the left- and right-handed elliptically polarized waves appears when the following condition is satisfied $\varepsilon_v=\mu_v=0$ at $k_z=0$. The strength of modes interaction within the found extreme states is also elucidated.

It is found out that the surface polaritons are of hybrid type, namely, they are EH and HE modes. It is revealed that providing a proper choice of the superlattice parameters, the conditions of degeneracy of the EH and HE surface polaritons can be obtained as a result of manifestation of the crossing effect. 

We argue that the discussed dispersion features of polaritons identified in the magnetic-semiconductor superlattice under study have a fundamental nature and are common to different types of waves and waveguide systems.

\appendix

\section{Derivation of dispersion equations for bulk and surface polaritons}
\label{sec:dispeq}

The electric and magnetic field vectors $\vec E$ and $\vec H$ can be represented in a general form as \cite{Tuz_JMMM_2016, Tuz_arXiv_2016}:
\begin{equation}
\vec P^{(j)}= \vec p^{(j)}\exp\left(ik_z z\right)\exp\left(\mp\kappa_j y\right), \label{eq:inc}
\end{equation}
where a time factor $\exp\left(-i \omega t\right)$ is also supposed and omitted; the sign `$-$' is related to the fields in the upper medium ($y>0$, $j=0$), while sign `$+$' is related to the fields in the composite medium ($y<0$, $j=1$), respectively.

From a pair of the curl Maxwell's equations $\nabla\times\vec E = i k_0 \vec B$ and $\nabla\times\vec H = -i k_0 \vec D$, where $k_0=\omega/c$ is the free space wavenumber, in a standard way we derive the following equation for the macroscopic field:
\begin{equation}
\nabla \times \nabla \times\vec P^{(j)} - k_0^2 \hat \varsigma^{(j)} \vec P^{(j)} = 0,
\label{eq:waveeq}
\end{equation}
where $\hat \varsigma$ is introduced as the product of $\hat \mu^{(j)}$ and $\hat \varepsilon^{(j)}$ made in the appropriate order.

For the upper medium ($j=0$), direct substitution of expression \eqref{eq:inc} with $\vec P^{(0)}$ and corresponding constitutive parameters ($\varsigma^{(0)}\equiv\varsigma_0 \hat I=\varepsilon_0\mu_0\hat I$, where $\hat I$ is the identity tensor) into equation (\ref{eq:waveeq}) gives us the relation with respect to $\kappa_0$:
\begin{equation}
\kappa_0^2 = k_z^2-k_0^2\varepsilon_0\mu_0. \label{eq:kappa0}
\end{equation}

For the gyroelectromagnetic medium ($j=1$), substitution of \eqref{eq:inc} into \eqref{eq:waveeq} with subsequent elimination of $P^{(1)}_y$ yields us the following set of two homogeneous algebraic equations for the rest two components of $\vec P^{(1)}$:
\begin{subequations}
\label{eq:systemAB}
\begin{align}
  \label{eq:systemABa}
  A_{xz}P^{(1)}_x + B_{xz}P^{(1)}_z&=0, \\
  \label{eq:systemABb}
  B_{zx}P^{(1)}_x + A_{zx}P^{(1)}_z&=0,
\end{align}
\end{subequations}
where coefficients $A_{xz}= k_z^2-\kappa^2-k_0^2\varsigma_{xx}$,  $A_{zx}= (k_0^2/ik_z \kappa)(k_z^2\varsigma_{zz}-\kappa^2\varsigma_{yy}-k_0^2\varsigma_{yy}\varsigma_{zz})$ and $B_{xz}= -(k_0^2\varsigma_{xy}/ik_z \kappa)(\kappa^2+k_0^2\varsigma_{zz})$ are functions of $\kappa$; $B_{zx}= -k_0^2\varsigma_{yx}$. 

In order for set of equations (\ref{eq:systemAB}) to have a nontrivial solution, the determinant of its coefficients must vanish. As a result we obtain a biquadratic equation with respect to $\kappa$
\begin{equation}
a\kappa^4+b\kappa^2+c=0, 
\label{eq:kappa1}
\end{equation}
where $a = \varsigma_{yy}$, $b = -k_z^2(\varsigma_{yy}+\varsigma_{zz})+k_0^2[\varsigma_{yy}(\varsigma_{xx}+\varsigma_{zz})-\varsigma_{xy}\varsigma_{yx}]$, $c = k_z^2\varsigma_{zz}[k_z^2-k_0^2(\varsigma_{xx}+\varsigma_{yy})] + k_0^4\varsigma_{zz}(\varsigma_{xx}\varsigma_{yy}-\varsigma_{xy}\varsigma_{yx})$, and whose solution is trivial. Dispersion relation \eqref{eq:DispBulk} for the \textit{bulk} polaritons is then obtained from \eqref{eq:kappa1} by putting $\kappa = 0$ inside it. 

In order to find the dispersion law of the surface polaritons from four roots of (\ref{eq:kappa1}) two physically correct roots ($\kappa_1$ and $\kappa_2$) must be selected. Then in accordance with \cite{Burstein_PhysRevB_1974}, we introduce the quantities $K_w$ ($w=1,2$) in the form:
\begin{subequations}
\label{eq:coeffKi}
\begin{align}
\label{eq:coeffKia}
P_x^{(1)}(\kappa_w) &= K_w A_{zx}(\kappa_w),  \\
\label{eq:coeffKib}
P_y^{(1)}(\kappa_w) &= K_w C(\kappa_w), \\
\label{eq:coeffKic}
P_z^{(1)}(\kappa_w) &= -K_w B_{zx}(\kappa_w),
\end{align}
\end{subequations}
where $C(\kappa_w)=-[(\kappa_w^2+k_0^2\varsigma_{zz})/ik_z\kappa_w]B_{zx}(\kappa_w)$, and these quantities $K_w$ need to be determined from the boundary conditions.

Taking into consideration that two appropriate roots $\kappa_1$ and $\kappa_2$ of equation (\ref{eq:kappa1}) are properly selected, the components of the field $\vec P^{(1)}$ can be rewritten as a linear superposition of two terms with respect to these roots:
\begin{subequations}
\label{eq:inckappa}
\begin{align}
\label{eq:inckappaa}
P_x^{(1)}&= \sum_{w=1,2}K_wA_{zx}(\kappa_w)\exp(\kappa_w y),  \\
\label{eq:inckappab}
P_y^{(1)}&= \sum_{w=1,2}K_wC(\kappa_w)\exp(\kappa_w y),  \\
\label{eq:inckappac}
P_z^{(1)}&= -\sum_{w=1,2}K_wB_{zx}(\kappa_w)\exp(\kappa_w y),  
\end{align}
\end{subequations}
where $y < 0$ and the factor $\exp\left[i (k_z z-\omega t)\right]$ is omitted.

Involving a pair of the divergent Maxwell's equations $\nabla \cdot \vec B=0$ and $\nabla \cdot \vec D=0$ and imposition the boundary conditions at the interface which require the continuity of the tangential components of $\vec E$ and $\vec H$ and the normal components of $\vec D$ and $\vec B$ (i.e. in our notations these components are $P_x$, $P_z$ and $Q_y$, respectively) give us the next set of four independent linear homogeneous algebraic equations with respect to the unknown amplitudes $K_1$, $K_2$ and $P^{(0)}_x$, $P^{(0)}_z$:  
\begin{subequations}
\label{eq:finalsystem}
\begin{align} 
\label{eq:finalsystema}
P_x^{(0)} &= \sum_{w=1,2}K_wA_{zx}(\kappa_w) , \\ 
\label{eq:finalsystemb}
P_z^{(0)} &= -\sum_{w=1,2}K_wB_{zx}(\kappa_w), \\
\label{eq:finalsystemc} 
\frac{\kappa_0 g_{zz}}{g_0}P_x^{(0)} &= -\sum_{w=1,2}\kappa_w K_w A_{zx}(\kappa_w), \\ 
\label{eq:finalsystemd}
\frac{\kappa_0^2-k_z^2}{\kappa_0g_0} g_v P^{(0)}_z &= \sum_{w=1,2}K_w Z(\kappa_w),
\end{align}
\end{subequations}
where $Z(\kappa_w) = (i k_z g_{xy}/g_{yy})A_{zx}(\kappa_w)+\kappa_w B_{zx}(\kappa_w)+i k_z C(\kappa_w)$, $g_v=g_{xx}+g_{xy}^2/g_{yy}$, and $g_{vv'}$ are elements of the tensor $\hat g$ which is substituted for the corresponding tensor of effective permittivity $\hat \varepsilon$ or effective permeability $\hat \mu$.

In order for set of equations (\ref{eq:finalsystem}) to have a nontrivial solution, the determinant of its coefficients must vanish. It gives us dispersion equation \eqref{eq:DispSurf} for \textit{surface} polaritons.

Finally, the amplitudes $K_1$ and $K_2$ can be found by solving the set of linear homogeneous equations \eqref{eq:finalsystem}. Amplitudes $P^{(0)}_x$ and $P^{(0)}_z$ then follow from relations \eqref{eq:finalsystema} and \eqref{eq:finalsystemb}.


\bibliographystyle{elsarticle-num}
\bibliography{hybrid.bib}






\end{document}